\documentclass[useAMS,usenatbib]{mn2e}
\usepackage{times}
\usepackage{mathptm}
\usepackage{graphicx}
\usepackage{aas_macros}

\newcommand{\msun}{\mbox{${\rm M}_{\odot}$ }}
\def\rev#1{{#1}}

\title[SN2010O: an explosion in an X-ray binary?]{The type Ib supernova 2010O: an explosion in a Wolf-Rayet X-ray binary?}

\author[Nelemans, Voss, Nielsen \& Roelofs]{Gijs Nelemans$^{1}$,\thanks{E-mail: nelemans@astro.ru.nl}  
 Rasmus Voss$^{2,3}$, Mikkel T.B. Nielsen$^{1}$ and Gijs Roelofs$^{4,5}$ 
  \\
 $^{1}$Radboud University Nijmegen, Department of Astrophysics/IMAPP,
  P.O. Box 9010, NL-6500 GL Nijmegen, The Netherlands\\
  $^{2}$Max Planck Institute of Extraterrestrial Physics,
  Giessenbachstrasse, 85748, Garching, Germany\\
  $^{3}$Excellence Cluster Universe, Technische Universit\"at
M\"unchen, Boltzmannstr.  2, D-85748, Garching, Germany\\ 
  $^{4}$Harvard-Smithsonian Center for Astrophysics, 60 Garden St, Cambridge, MA 02138, USA\\
 $^{5}$Netherlands Bureau for Economic Policy Analysis, Van Stolkweg 14, 2585 JR The Hague, The Netherlands\\ 
}

\begin{document}

\date{Accepted ... Received \today}

\pagerange{\pageref{firstpage}--\pageref{lastpage}} \pubyear{2010}

\maketitle

\label{firstpage}

\begin{abstract}
The type Ib supernova 2010O was recently discovered in the interacting
starburst galaxy Arp 299. We present an analysis of two archival
\emph{Chandra} X-ray observations of Arp 299, taken before the
explosion and show that there is a transient X-ray source at a
position consistent with the supernova. Due to the diffuse emission,
the background is difficult to estimate. We estimate the flux of the
transient from the difference of the two X-ray images and conclude
that the transient can be described by a \rev{0.225 keV black body
with a luminosity of $2.5\pm0.7 \times 10^{39}$ erg s$^{-1}$} for a
distance of 41 Mpc. These properties put the transient in between the
Galactic black hole binary XTE J1550-564 and the ultra-luminous X-ray
binaries NGC 1313 X-1 and X-2. The high level of X-ray variability
associated with the active starburst makes it impossible to rule out a
chance alignment. If the source is associated with the supernova, it
suggests SN2010O is the explosion of the second star in a Wolf-Rayet
X-ray binary, such as Cyg X-3, IC 10 X-1 and NGC 300 X-1.
\end{abstract}


\section{Introduction}\label{introduction}

Type Ib supernovae lack hydrogen in their spectrum, but show helium lines,
while Ic supernovae also lack helium. They are regarded as the core-collapse
explosions of massive stars that have lost their hydrogen envelopes and thus
have become Wolf-Rayet stars \citep{1986ApJ...306L..77G}. These Wolf-Rayet
stars can originate from the most massive stars that can remove their hydrogen
envelope on and just after the main sequence in strong stellar winds, or from
lower-mass stars via a binary interaction that removes the hydrogen envelope
\citep[e.g.][]{1992ApJ...391..246P,1995PhR...256..173N}.

In the last decade a new way of linking supernovae with their progenitors has
become available. The growing archive of high-resolution images has made it
possible to detect the progenitors of type II (hydrogen rich) supernovae in
pre-explosion optical images \citep[see][for a
  review]{2009ARA&A..47...63S}. No progenitors of type Ib or Ic supernovae
have been found, despite deep pre-explosion images for 10 of them. However,
the relative frequency of Ib/Ic to type II supernovae of around 0.4 suggests
that binary interactions play an important role, as for a standard initial
mass function, there are not enough very massive stars that could lose their
envelope via a stellar wind \citep{2009ARA&A..47...63S}.

An alternative method for directly detecting supernova progenitors is
to use archival images in other wave bands, such as X-ray data. We
started a program to search for type Ia supernova progenitors in
\emph{Chandra} X-ray data and have found one likely progenitor and
derived five upper limits \citep{vn08,rbv+08,nvr+08}. For type Ia
supernova progenitors the argument to look for X-ray progenitors is
the suggestion that supersoft X-ray sources may produce type Ia
supernovae, based on the accreting white dwarf model
\citep{1973ApJ...186.1007W,nom82}. For type Ib and Ic supernovae, the
binary progenitor scenario suggests the possibility that the Ib or Ic
explosion is the second supernova in the binary and thus before the
explosion may have been part of a high-mass X-ray binary. Indeed, the
relative frequency estimates of \citet[][their figure
16]{1992ApJ...391..246P} suggest that the likelihood for a Ib to be
the second explosion in the binary is at least as high as it being the
first one. It is therefore useful to constrain the X-ray properties of
the direct progenitors of Ib and Ic supernovae.

Even more tantalising evidence for such scenario comes from the known
Wolf-Rayet star in the X-ray binary Cyg X-3 \citep{1996A&A...314..521V} and
the recent discovery of two X-ray binaries in which a (massive) Wolf-Rayet
star orbits a black hole, IC 10 X-1 and NGC 300 X-1
\citep{2004A&A...414L..45C,2004ApJ...601L..67B,2007ApJ...669L..21P,2007A&A...461L...9C}.
Given the short life times of massive Wolf-Rayet stars, a Ib or Ic supernova
explosion is expected within next few million years in these systems.

In this paper we report the discovery of a transient soft X-ray source
at the position of the recent Ib supernovae 2010O that exploded in the
interacting starburst galaxy Arp 299 (\rev{IC 694}/NGC 3690). In
Sect.~\ref{2010O} we discuss supernova 2010O and its host galaxy, in
Sect.~\ref{obs} the X-ray, optical and radio observations that we used
and in Sect.~\ref{results} the results of the \emph{Chandra}
analysis. In Sect.~\ref{discussion} we discuss the implications of the
finding, the possibility of a chance alignment and the scope for
future work.

\section{Supernova 2010O in Arp 299}\label{2010O}

%\begin{figure}
%\includegraphics[width=\columnwidth,clip]{new.jpg}
%\caption[]{Hubble Space Telescope Heritage project composite image of
%  Arp 299, indicating the approximate position of SN 2010O with the arrow.}
%\label{fig:arp}
%\end{figure}

SN20100 was discovered on Jan 24, 2010 in the course of the Puckett
Observatory Supernova Search \citep{2010CBET.2144....2N} and a
spectrum taken with the Nordic Optical telescope on Jan 28, 2010
showed it to be a type Ib supernova \citep{2010CBET.2149....1M}. The
supernova resides in the Eastern part of the interacting galaxy Arp
299, which is often referred to as IC 694 in the literature,
designating the Western part as NGC 3690. However, \rev{as this
convention is somewhat confusing\footnote{\citet{1999AJ....118..162H}
discuss that it is likely, although not conclusively so that Swift
referred to the compact dwarf to the northwest of Arp 299 for IC
694. All supernovae in \rev{either Eastern or Western part of} Arp 299
(including SN 2010O) are classified as ``in NGC 3690'' by the Central
Bureau for Astronomical Telegrams of the IAU.}, we use the name Arp
299 to refer to the whole system and use Eastern and Western part to
identify the two galaxies.}

Arp 299, at a distance of about 41 Mpc, is a pair of interacting
galaxies giving rise to a spectacular starburst
\citep[e.g.][]{1983ApJ...267..551G} and has in the last 20 years
produced seven supernovae: 1992bu (type II), 1993G (type II), 1998T
(type Ib), 1999D (type II), 2005U (type II or IIb), 2010O (type Ib)
and SN2010P (type unknown) or eight if including the possible radio
supernova reported by \citet{1990IAUC.4988....1H}. The star formation
rate is estimated to be \rev{ranging from 22--140} \msun yr$^{-1}$
\citep{2000ApJ...532..845A} and \emph{Chandra} X-ray images reveal
extended emission and at least 18 discrete sources of which one is an
obscured AGN in the Western part of the interacting galaxy
\citep{2003ApJ...594L..31Z}. Radio observations show several discrete
sources, as well as extended emission in the nuclei of both parts
which have been resolved in 30 individual sources using VLBI
\citep[e.g.][]{2007ApJS..173..185G,2004ApJ...611..186N,2009AJ....138.1529U,2009A&A...507L..17P},
likely all young supernova remnants.

Clearly Arp 299 is a very complex stellar system with tremendous
activity, making association of different sources with each other
difficult. Nevertheless, we will analyse the pre-supernova X-ray data
of Arp 299 to search for possible emission of the progenitor of 2010O.

\section{Observations}\label{obs}

\begin{table}
\caption[]{List of pre-SN \emph{Chandra} observations. \rev{The pointing
  offset between the source and the center of the \emph{Chandra} field
  is given in the last column.}}
\label{tab:obs}
\begin{tabular}{llllrl}
Instrument & ID & date & exp. time & offset \\ \hline
ACIS-I & 1641 & 2001-07-13 & 24850  & 1.2'  \\
ACIS-S & 6227 & 2005-02-14 & 10320   & 0.4' \\\hline
\end{tabular}
\end{table}

There are \emph{Chandra} images at two epochs before the supernova
(taken in 2001 and 2005), 25 ks ACIS-I and 10 ks ACIS-S. The 2001 data
is published in \citet{2003ApJ...594L..31Z}, while the 2005 DDT
observation (aimed at detecting an X-ray flare from the AGN in the
\rev{Western} part) has not been published.  In table~\ref{tab:obs} we show
a log of the observations.

We checked the alignment of the two X-ray images by comparing the
positions of seven isolated sources in the two images resulting in a
difference of \rev{$\Delta x=$0.02$\pm$0.05; $\Delta
y=-$0.11$\pm$0.07} pixels. With a pixel scale of 0$\farcs$491 this
translates into a shift of 0$\farcs$05 with marginal significance. We
therefore chose not to correct for this possible shift.

We checked the absolute astrometric accuracy of the X-ray images by
comparing them to USNO B1.0 and 2MASS positions, but there are only
two matches and they are off-axis in the X-ray images so we can only
conclude that they agree to within 1$\farcs$ A more useful check comes
from the positions of several radio sources found in Arp 299
\citep{2004ApJ...611..186N}. Two of those sources (B1 and D) match
with X-ray sources and we find agreement to within half a pixel, with
a hint of a small \rev{(0$\farcs$25)} offset of the X-ray image to the
North.

We also checked the astrometry of the optical discovery image
\citep{2010CBET.2144....2N} and the image taken by Joseph
Brimacombe\footnote{http://www.flickr.com/photos/43846774@N02/4308201580/} and
kindly provided to us.  Comparison of the positions of 15 USNO B1.0 and 2MASS
stars gives an RMS scatter of 0$\farcs$25. Together with the uncertainties in
the USNO B1.0 and 2MASS positions and the uncertainty of about 0$\farcs$4 in
the position of the supernova on top of the galaxy, we estimate the accuracy of
the optical position at 0$\farcs$5. The best fit position of the supernova in
these images is RA = 11:28:33.811, DEC=+58:33:51.75, 0$\farcs$4 away from the
reported discovery position.

\begin{figure}
\includegraphics[width=\columnwidth,clip]{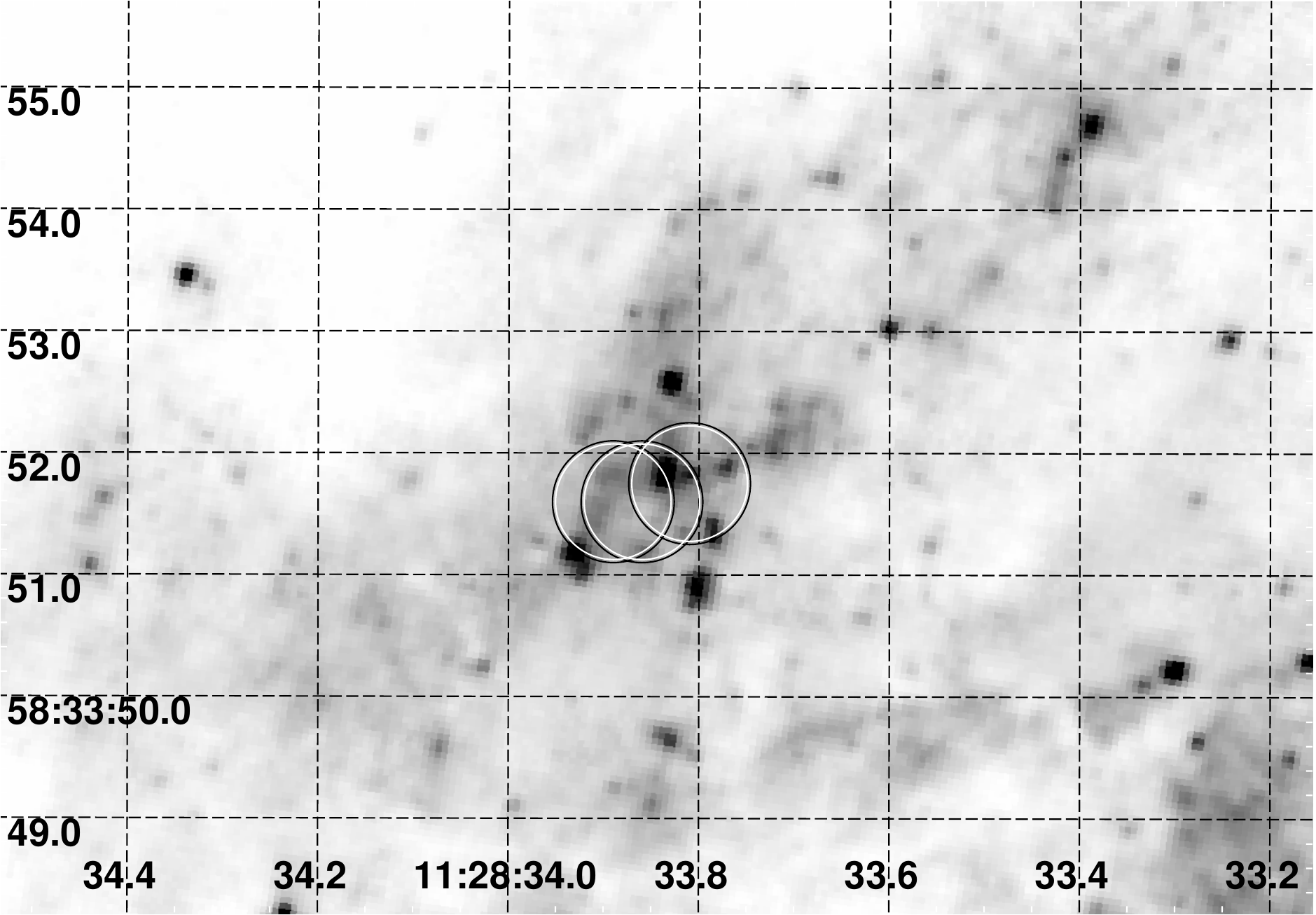}
\caption[]{\emph{Hubble Space Telescope} ACS F814W image of the region
  around SN2010O. The three circles indicate \rev{(right to left): our
  best optical position, the reported discovery position (consistent
  with the position shown in \citet{2010ATel.2422....1B}), and the
  best position of the X-ray transient (Sect.~\ref{results})}. The
  radius of the circles is 0$\farcs$5.}
\label{fig:ACS}
\end{figure}

\begin{figure*}
\includegraphics[width=0.4\textwidth,clip]{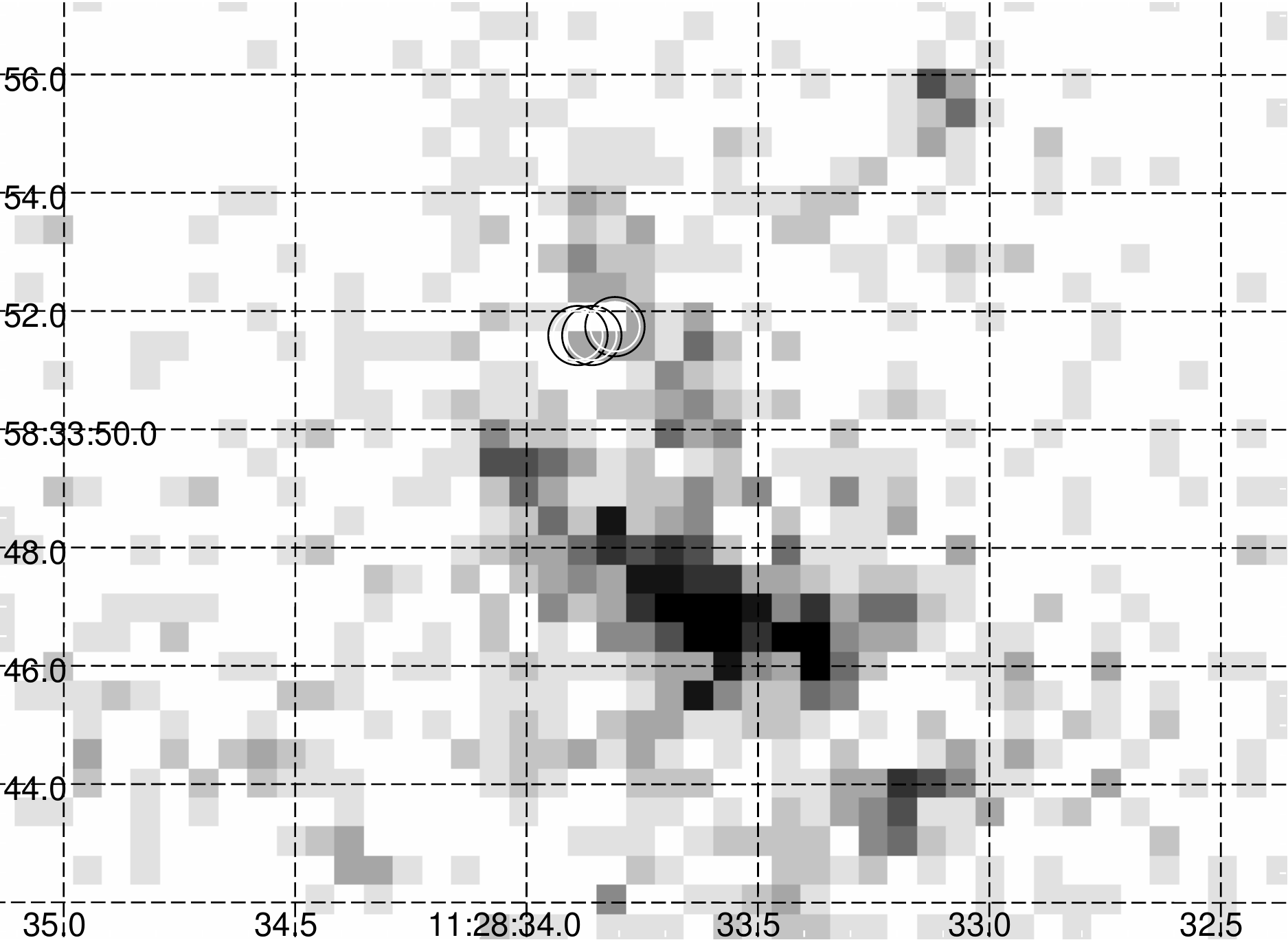}
\includegraphics[width=0.4\textwidth,clip]{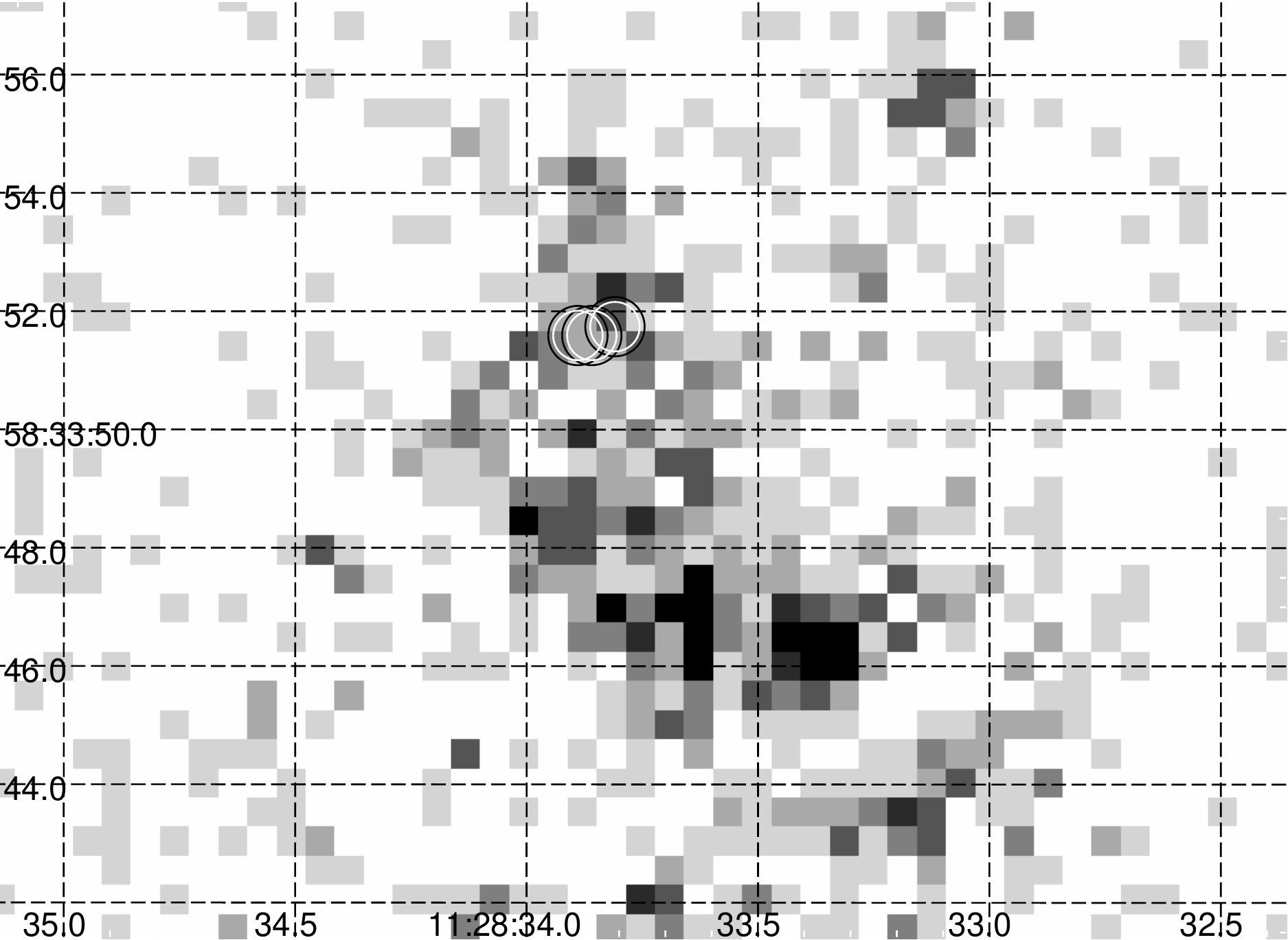}

\includegraphics[width=0.4\textwidth,clip]{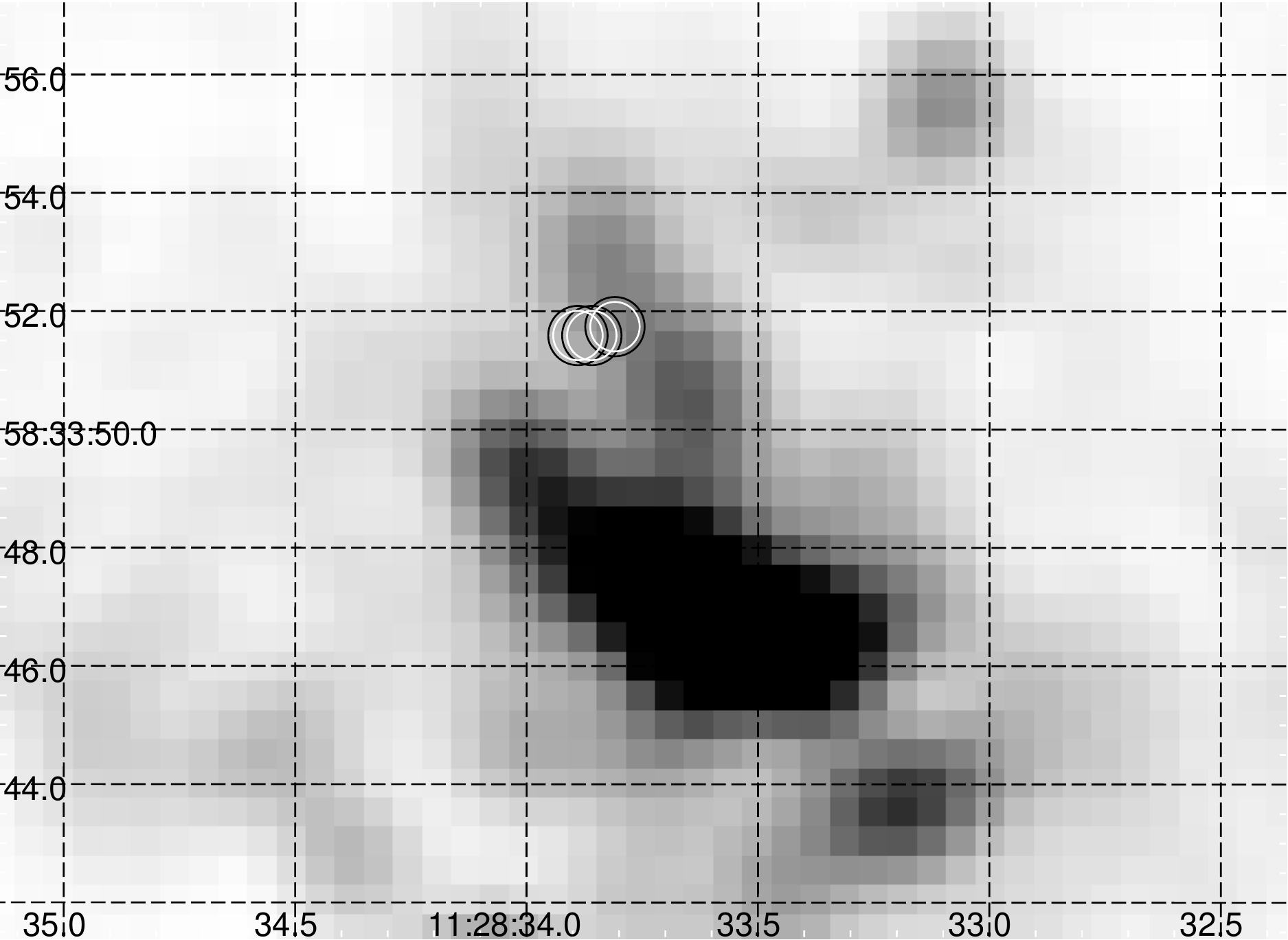}
\includegraphics[width=0.4\textwidth,clip]{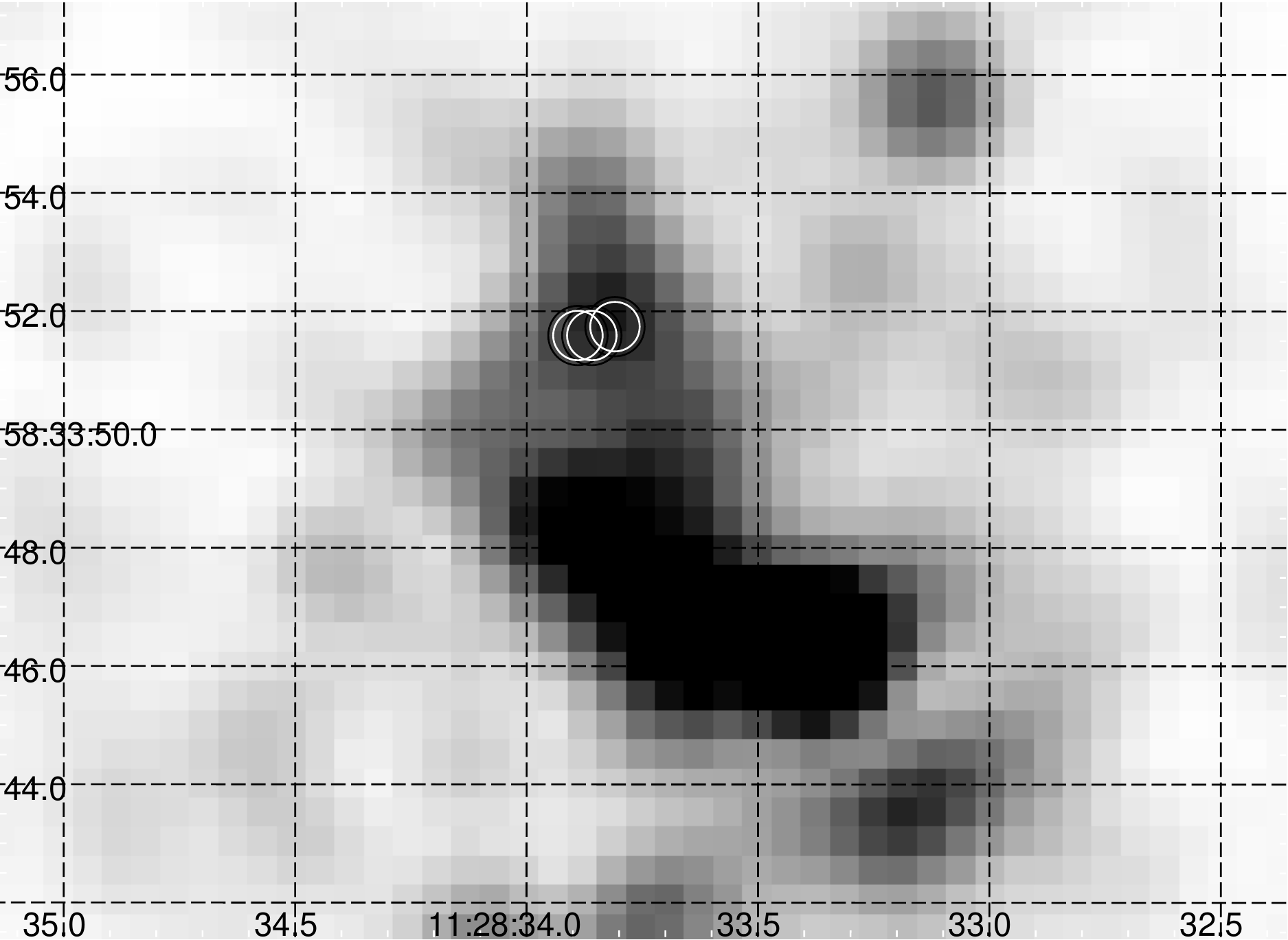}

\caption[]{\emph{Chandra} images of the region around 2010O before the
  explosion. Top: observation \rev{1641} (left) and observation \rev{6227}
  (right). Bottom: same images Gaussian smoothed by 3 pixels. \rev{Our best
    optical position and the reported optical position of 2010O are indicated
    as the right two 0$\farcs$5 radius circles. The SExtractor position of the
    source in the difference image (Fig.~\ref{fig:diff}) is the left
  most circle.}}
\label{fig:X}
\end{figure*}

\citet{2010ATel.2422....1B} report observations with the WIYN 0.9m
telescope that have been registered to the SDSS image, which then was
registered to the archival \emph{Hubble Space Telescope} ACS
images. They report that the position of the supernova is consistent
with a bright blue, slightly resolved object, likely a young star
cluster. Fig.~\ref{fig:ACS} shows the region of the supernova in the
ACS image, with the different reported positions (including that of
the X-ray source discussed below) indicated. The optical positions
agree to within the uncertainties of 0$\farcs$5. We have registered
the ACS image with the single radio source D from
\citet{2004ApJ...611..186N}, leaving some uncertainty in the absolute
calibration of the circles relative to the image.

\section{Results of the \emph{Chandra} analysis}\label{results}

In Fig.~\ref{fig:X} we show the two \emph{Chandra} images of the
region. WAVDETECT detects a source with \rev{formal significance of
9.2 sigma} in observation 6227 at position RA=11:28:33.813,
DEC=+58:33:51.95, i.e. 0$\farcs$15 from our best optical position and
0$\farcs$5 from the reported discovery position. \rev{However, the
X-ray source is extended and includes a ridge of emission that is also
present in observation 1641 (see Fig.~\ref{fig:X})}. There is no
detection of a source in observation 1641. However, at the position of
the supernova there is clear diffuse emission (see Fig.~\ref{fig:X}),
complicating the analysis and artificially increasing the significance
of the detection in observation 6227. \rev{We therefore subtracted
image 6227 from (a scaled version of, see below) image 1641 and show
the result in Fig.~\ref{fig:diff}. There is a point like source that
suggest there is a transient X-ray source at a position consistent
with the position of the supernova. After the variable
central source it is the most significant source in the
difference image. We ran SExtractor \citep{1996A&AS..117..393B} on the
difference image and clearly detect the transient source at position
RA=11:28:33.89, DEC=+58:33:51.6.}

\begin{figure}
\includegraphics[width=\columnwidth,clip]{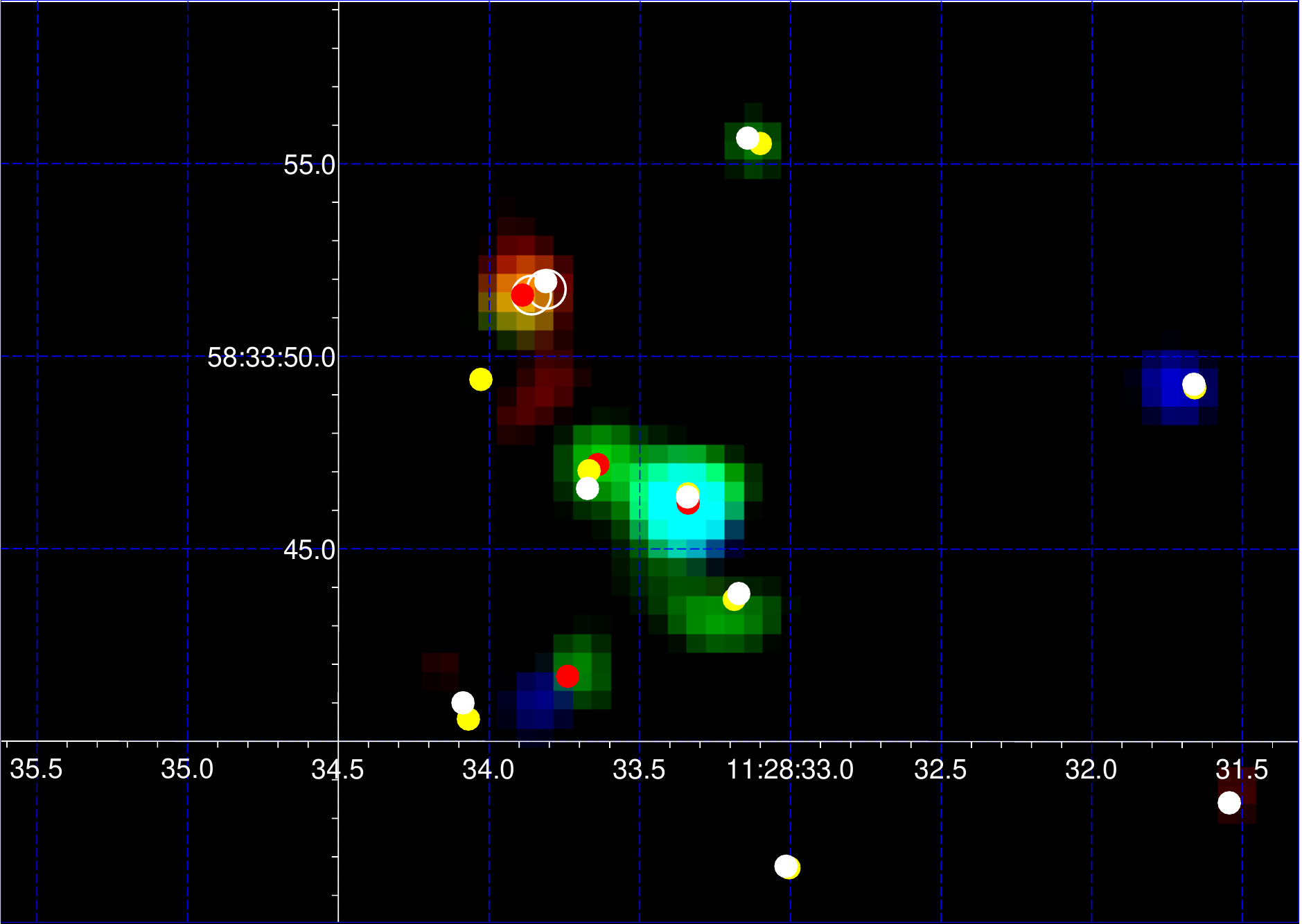}
\caption[]{Colour image of the \emph{difference} between
  \emph{Chandra} observation 6227 and the scaled 1641 observation. In
  red the soft (0.3-1 keV) in green the medium (1-2 keV) and in blue
  the hard (2-8 keV) X-ray image.  \rev{The positions of sources found
  with WAVDETECT are indicated with yellow symbols for sources in
  observation 1641 and with white symbols for sources in observation
  6227. The four sources found with SExtractor in the difference image
  (see text) are indicated with red symbols. The reported position of
  2010O and our best optical position are indicated as the two
  white 0$\farcs$5 radius circles.} }
\label{fig:diff}
\end{figure}

In the following we proceed on the assumption that the local
background has not changed significantly, i.e. that we can use
observation 1641 as an estimate of the background. One should keep in
mind that it is quite possible that the source is also present in the
earlier observation. Because the two observations were taken with
different chips, we first determine the relative sensitivity to the
diffuse emission for the two observations in three different bands
(0.3-1, 1-2 and 2-8 keV) assuming:\\ a) an absorbed black body of
temperature 1 keV, taking the absorption as $N_{\rm h} = 10^{21}$
\rev{cm$^{-2}$}, based on the reported reddening of the supernova
spectrum of $A_{\rm V} = 0.65$ \citep{2010CBET.2149....1M}.\\ b) an
absorbed power-law with index 2 and \\c) a power-law with index 2 and
only the expected foreground absorption ($N_{\rm h} = 10^{20}$
\rev{cm$^{-2}$}).

\begin{table}
\caption[]{Relative sensitivity and counts in the diffuse emission around the
  position of the transient source in the two observations, expressed as ratio
  of observation 1641 to 6227.}
\label{tab:sens}
\begin{tabular}{lllll}
Band & high ab. BB & high ab. PL & low ab. PL & counts \\ \hline
Low (0.3-1 keV) & 1.07 & 0.91 & 0.83  & 1.18   \\
Middle (1-2 keV) & 1.61 & 1.61 &  1.61 & 1.63 \\
High (2-8 keV) & 1.82 & 1.80 & 1.80 & 1.70  \\
Full & 1.69 & 1.42 & 1.30  & 1.45  \\\hline
\end{tabular}
\end{table}

We compare these to the detected counts in the two observations in a
region around the transient X-ray source with radius 100 pixels. The
results are shown in table~\ref{tab:sens}. We conclude that the ratios
measured in the two observations agree roughly with the expectations
(in particular for the absorbed black body). We thus divide the counts
in observation 1641 by these ratios and assume that gives a reasonable
estimate of the diffuse background. \rev{Within a 1$\farcs$0 radius
(which should contain 90 per cent of the flux for a source around 0.4
arc min off-axis) we find 32 photons in observation 6227, while the
background estimate based on the scaled observation 1641, is 10.0. For
standard Poisson statistics the likelihood of detecting 32 photons or
more for an expectation of 10.0 is $7 \times 10^{-9}$, corresponding
to more than 5.5 sigma. However, the background value is determined
from only a small region in observation 1641, using 14 counts, and
thus has a relatively large intrinsic uncertainty. We therefore used a
small Monte Carlo simulation, where we generated 25 million
backgrounds, taking the (Poisson) uncertainty on the 14 counts in the
background determination into account. The result is that the
probability of finding 32 photons or more increases to $3 \times
10^{-5}$ or about 4 sigma.}

% Monte carlo: 750 out of 25e6 have N > 31

\rev{Translating the counts in the difference image to X-ray flux \rev{using
    the black body model}, we obtain an unabsorbed flux of 9.1 $\times
 10^{-15}$ erg cm$^{-2}$ s$^{-1}$. }

\rev{The counts in the different bands \rev{(0.3-1, 1-2 and 2-8 keV)} for the
  difference between the scaled observation 1641 and 6227 are 8.9, 11.3 and
  1.8 respectively. This is softer than the spectra used above to determine
  the overall sensitivity difference. Indeed, the counts are consistent with a
  black body of temperature of 0.225 keV}

We thus repeated the calculation of the fluxes with a \rev{0.225 keV
black body and find an absorbed flux of 8.6 $\times 10^{-15}$ erg
cm$^{-2}$ s$^{-1}$ and an unabsorbed flux of 1.2 $\times 10^{-14}$ erg
cm$^{-2}$ s$^{-1}$. For a distance of 41 Mpc this corresponds to an
unabsorbed luminosity of 2.5 $\times 10^{39}$ erg s$^{-1}$.} A Monte
Carlo calculation generating the background as described above and
adding sources of different luminosity provides the 1 sigma
uncertainties on this \rev{luminosity: $L_{\rm X} = 2.5\pm0.7 \times
10^{39}$ erg s$^{-1}$.}

% statistics done with 2010O_new.py --> 2010O_new.dat
% cumulative histogram between 1 and 60. 
% median 23, 1 sigma points at
% y = 0.158 and 0.842: 17 and 29, i.e. relative error 0.26

\section{Discussion and conclusion}\label{discussion}

If the X-ray transient found in the second \emph{Chandra} image is associated
with the SN2010O (see below for a discussion of the possibility of a chance
alignment), it suggests that this is the explosion of the second star in a
binary system. Before the supernova explosion the system must have consisted
of a Wolf-Rayet star orbiting a compact object. We can compare the X-ray
properties of the transients to the outbursts seen in Galactic black-hole
X-ray binaries \citep[see][for an overview]{mr04}. The peak luminosity is
higher than those of the Galactic sources, but these are particularly poorly
known, as the distances to these systems are not well determined
\citep[see][]{jn04}. The outburst luminosity is below the Eddington limit of a
10 \msun black hole, if we assume pure helium accretion. The spectrum is much
softer than the typical 1 keV found in Galactic black hole binaries. However,
the very soft spectrum and high luminosity fit very well in between the
highest luminosity point of the Galactic X-ray binary XTE J1550-564 and the
ultra-luminous X-ray binaries NGC 1313 X-1 and X-2 \citep[see figure 7
  of][]{2007Ap&SS.311..213S}.

The transient thus could be a relative of the nearby Wolf-Rayet X-ray
binaries Cyg X-3 \citep{1996A&A...314..521V}, IC 10 X-1
\citep{2004A&A...414L..45C,2004ApJ...601L..67B} and NGC 300 X-1
\citep{2007A&A...461L...9C} and the soft spectrum suggests a (massive)
black hole accretor. However, all of these seem to be fairly
persistent X-ray sources, with luminosities well below the Eddington
limit. The transient thus would represent the equivalent of the
outburst in soft X-ray transients, in a Wolf-Rayet X-ray binary.

With a distance modulus of 33 and and a reported visual magnitude of
15.8 around the peak \citep{2010CBET.2144....2N}, together with the
extinction of 0.65 as derived from the spectrum
\citep{2010CBET.2149....1M}, the absolute magnitude of SN2010O is
$-17.9$. This is well within the large range of absolute magnitudes of
type Ib supernovae \citep{2006AJ....131.2233R}.  Unfortunately, there
are no known correlations of Ib supernova explosion features and the
properties of the progenitor Wolf-Rayet stars.

The whole discussion above depends on the association of the X-ray
transient with the supernova. With WAVDETECT 22 unique sources are
found within half an arc min from the position of the supernova. If we
assume we would consider sources within 1$\arcsec$ as a possible match
that would give a chance alignment probability of 2 per cent. Ten of
these sources are actually within 15$\arcsec$ and the difference image
in Fig.~\ref{fig:diff} shows \rev{two sources not detected by
WAVDETECT if we include the transient}. We therefore conclude that
there is a 5 per cent probability of a chance alignment.  However, the
close match in position and general consistency of the X-ray
properties with what could be expected from an X-ray binary in which
the second star exploded as a Ib supernova, make the connection
plausible. \rev{The possible association of the supernova with the
cluster seen in the \emph{HST} ACS observations
\citep{2010ATel.2422....1B} also allows for the possibility that the
X-ray source is coincident with the position of the SN, but is
actually an unrelated X-ray binary in the cluster. However this is not
very likely, as there are only 18 X-ray sources found in Arp 299, and
many more star clusters, and a study by \citet{2004MNRAS.348L..28K}
shows that in starburst galaxies the clusters have only rarely an
X-ray source right on top of them}.

We therefore conclude that the observations presented in this paper suggest
the exciting possibility that supernova 2010O was the explosion of a massive
Wolf-Rayet star that was part of a X-ray binary containing a black hole.

Further study, in particular for other supernovae with pre-supernova
\emph{Chandra} data, will be needed to address the question of how often type
Ib and Ic supernovae explode in X-ray binaries. We are in the process of
analysing the earlier type Ib and Ic supernovae for which there is archival
\emph{Chandra} data (Nielsen et al. in preparation).

\section*{Acknowledgments}

We thank Joseph Brimacombe for providing us the fits files of his
observations, Peter Jonker for discussions on the properties of black
hole binaries and the referee for helpful comments. We thank the
Central Bureau for Astronomical Telegrams for providing a list of
supernovae. This research has made use of data obtained from the
Chandra Data Archive and software provided by the Chandra X-ray Center
in the application package CIAO. GN and MN are supported by the
Netherlands Organisation of Scientific Research. This research was
supported by the DFG cluster of excellence ``Origin and Structure of
the Universe''.

\bibliography{journals,binaries} \bibliographystyle{mn2e}

\label{lastpage}

\end{document}